\documentclass[%
,twocolumn%
 ,secnumarabic%
,amssymb, amsmath,nobibnotes, aps, prl]{revtex4}
\usepackage[]{graphicx}
\usepackage{tabularx}
\usepackage{bm}%
\expandafter\ifx\csname package@font\endcsname\relax\else
 \expandafter\expandafter
 \expandafter\usepackage
 \expandafter\expandafter
 \expandafter{\csname package@font\endcsname}%
\fi

\begin{document}

\title{Geometry in transition in four dimensions:\\ A model of emergent geometry in the early universe and dark energy}

\author{Badis Ydri, Ramda Khaled,  Rouag Ahlam}
\affiliation{
Department of Physics, Faculty of Sciences, Badji Mokhtar Annaba University,\\
 Annaba, Algeria.
}

\begin{abstract}
We study a six matrix model with global $SO(3)\times SO(3)$ symmetry containing
at most quartic powers of the matrices.  This theory exhibits a phase transition from a geometrical phase  at low temperature to a Yang-Mills  matrix phase with no background geometrical structure at high temperature. This is an exotic phase transition in the same universality class as the three matrix model but with important differences. The geometrical phase is determined dynamically, as the system cools, and is given by a fuzzy four-sphere background ${\bf S}^2_N\times{\bf S}^2_N$,  with an Abelian gauge field which is very weakly coupled to two normal scalar fields playing the role of dark energy.
\end{abstract}

\maketitle

{\bf Introduction:}
The notion of geometry as an emergent concept is not new. See 
for example \cite{Bombelli:1987aa,Ambjorn:2007jv} for an inspiring discussion, along the lines of causal sets and lattice dynamical triangulation respectively,  and 
\cite{Seiberg:2006wf,Di Francesco:1993nw} for 
some other recent ideas from strings and random matrix theory. Another powerful approach is the idea of emergent noncommutative geometry \cite{connes} from IKKT Yang-Mills matrix models \cite{Ishibashi:1996xs}. 

We examine such a phenomenon in the context of 
noncommutative geometry  emerging from matrix models
by studying a surprisingly rich six matrix model proposed in \cite{CastroVillarreal:2005uu,DelgadilloBlando:2006dp}. This is a theory with two independent parameters, the gauge coupling $g$ and the mass deformation $M$, where the particular value $M=1/2$ was considered in  \cite{Behr:2005wp}. This model is a generalization of the three matrix model studied in 
\cite{Azuma:2004zq,CastroVillarreal:2004vh,O'Connor:2006wv}. 
The matrix geometry which emerges here is also interesting because 
it provides an alternative setting for the regularization of (noncommutative) field theories
\cite{Ydri:2001pv,O'Connor:2003aj,Balachandran:2005ew,Grosse:1996mz}, and also as 
the configurations of $D0$ branes in particular string theories, namely in the large $k$ limit of a boundary 
Wess-Zumino-Novikov-Witten model \cite{Myers:1999ps,Alekseev:2000fd,Knizhnik:1988ak}. Here, however, 
the geometry emerges as the system cools, much as a Bose condensate or superfluid
emerges as a collective phenomenon at low temperatures. And there is no background 
geometry in the high temperature phase. 

We consider the most general single trace Euclidean action functional for a 
six matrix model invariant under global $SO(3)\times SO(3)$ transformations containing no higher than 
the fourth power of the matrices. 
We find that 
generically the model has three phases. The first distinct phase is a geometrical one whereas the other distinct phase is a matrix phase. The third phase is a crossover phase which appears between the geometrical and the matrix phases for large values of $M$. 
The geometrical phase appears at low temperature (weak coupling). 
Small fluctuations in this phase are those of a Yang-Mills
 theory coupled to two scalar fields around 
a ground  state corresponding to the Cartesian product of two round fuzzy spheres \cite{HoppeMadore}, viz ${\bf S}^2_N\times{\bf S}^2_N$. The gauge group is Abelian. In the strict large $N$ limit the geometry becomes classical. As the temperature is increased the geometry undergoes a transition. In the matrix phase there is no background spacetime geometry and the fluctuations
are those of the matrix entries around zero. In this high temperature (strong coupling) phase the model is essentially
a zero dimensional reduction of $6-$dimensional Yang-Mills theory.

{\bf The model:}
Let $X_a$ and $Y_a$, $a=1,2,3$, be six $N{\times}N$
Hermitian matrices and let us consider the action
\begin{eqnarray}
\label{main2} 
S&=&S_1+S_2+S_{12}\nonumber\\
S_1&=&N\bigg[-\frac{1}{4}Tr[X_a,X_b]^2+\frac{2i\alpha}{3}{\epsilon}_{abc}TrX_aX_bX_c\nonumber\\
&+&\beta TrX_a^2+M Tr(X_a^2)^2\bigg]\nonumber\\
S_2&=&N\bigg[-\frac{1}{4}Tr[Y_a,Y_b]^2
+\frac{2i\alpha}{3}{\epsilon}_{abc}TrY_aY_bY_c\nonumber\\
&+&\beta TrY_a^2+M Tr(Y_a^2)^2\bigg]\nonumber\\
&&S_{12}=N\bigg[-\frac{1}{2}Tr[X_a,Y_b]^2\bigg].
\end{eqnarray}
The gauge coupling constant $\tilde{\alpha}^4=\alpha^4N^2=\beta$ plays the r{o}le of inverse temperature, the mass parameter $M$ controls the stability of the geometry, and we
fix $N=N_0^2$, $c_2^0=(N_0^2-1)/4$ and $\beta=-\alpha^2\mu~,~\mu=2(4c_{2}^0M-1)/9$ in this study.

The absolute minimum of the action is given by $X_a=\alpha\phi_0 L_a\otimes {\bf 1}_{N_0}$ and $Y_a=\alpha\phi_0 {\bf 1}_{N_0}\otimes L_a$ with $\phi_0=2/3$ and $L_a$ are the generators of $SU(2)$ in the irreducible representation of size $N_0$. 
Expanding around this configuration, with $X_a=\alpha\phi_0(L_a\otimes{\bf 1}+A_a)$ and  $Y_a=\alpha\phi_0({\bf 1}\otimes L_a+B_a)$, yields a noncommutative Yang-Mills 
action with gauge coupling $g^2=1/\tilde{\alpha}^4$. This theory includes two adjoint scalar fields, which are the components of the gauge field normal to the two spheres, given by

\begin{equation}
\Phi^1=\frac{1}{2}(x_aA_a+A_ax_a+\frac{A_a^2}{\sqrt{c_2^0}})~,~\Phi^2=\frac{1}{2}(y_aB_a+B_ay_a+\frac{B_a^2}{\sqrt{c_2^0}}).
\end{equation}
In the large $N$ limit, taken with $\tilde{\alpha}$ and $m^2=NM/2$ held fixed,
the action for small fluctuations becomes that of a $U(1)$ gauge field very weakly coupled to the above
 two scalar fields defined on a background commutative four-sphere ${\bf S}^2\otimes{\bf S}^2$. 
For large $m^2$ the two
scalar fields are simply not excited. 

One can see the background geometry as that of a fuzzy four-sphere with coordinates $x_a=L_a\otimes {\bf 1}_{N_0}/\sqrt{c_2^0}$ and $y_a={\bf 1}_{N_0}\otimes L_a/\sqrt{c_2^0}$ satisfying

\begin{eqnarray}
&&x_1^2+x_2^2+x_3^2=1~,~
[x_a,x_b]=\frac{i}{\sqrt{c_2^0}}{\epsilon}_{abc}x_c\nonumber\\
&&y_1^2+y_2^2+y_3^2=1~,~
[y_a,y_b]=\frac{i}{\sqrt{c_2^0}}{\epsilon}_{abc}y_c,
\end{eqnarray}
and
\begin{eqnarray}
[x_a,y_b]=0.
\end{eqnarray}
The algebra generated by products of the $x_a$ and $y_a$ is the algebra of all $N\times N$ 
matrices with complex coefficients. The geometry enters through 
the Laplacian \cite{O'Connor:2003aj} 
\begin{eqnarray}
\hat{\cal L}^2{\bf\cdot}=[L_a,[L_a,{\bf\cdot}]]\otimes{\bf 1}_{N_0}+{\bf 1}_{N_0}\otimes [L_a,[L_a,{\bf\cdot}]],
\end{eqnarray} 
which has the same spectrum as the round Laplacian on the commutative four-sphere ${\bf S}^2\times{\bf S}^2$,
but cut off on each sphere at a maximum angular momentum $L=N_0-1$. The fluctuations
of the scalar fields have this Laplacian as kinetic term.

The ground state is found by considering the configuration $X_a=\alpha{\phi}L_a\otimes{\bf 1}_{N_0}$ and $Y_a={\bf 1}_{N_0}\otimes \alpha\phi_0 L_a$ where
$\phi$ plays the role of the radius of the spheres defined by
 \begin{eqnarray}
&&{\cal R}^2=\frac{1}{N} Tr X_a^2~{\rm or}~{\cal R}^2=\frac{1}{N}TrY_a^2.\label{radiuss}
\end{eqnarray}
The radius $R$ was defined in \cite{DelgadilloBlando:2007vx} by the formula 
\begin{eqnarray}
\frac{1}{R}=\frac{1}{\phi_0^2 \tilde{\alpha}^2 c_2^0}Tr X_a^2.
\end{eqnarray}
The effective potential \cite{CastroVillarreal:2004vh,Azuma:2004ie,Ydri_in_preparation} 
obtained by integrating out fluctuations 
around the  ${\bf S}^2\times{\bf S}^2$ background is given,
in the large $N$ limit, by
\begin{eqnarray}
\frac{V}{2N^2}
&=&\tilde{\alpha}_0^4\bigg[\frac{\phi^4}{4}-\frac{{\phi}^3}{3}+m^2\frac{\phi^4}{4}-\mu\frac{\phi^2}{2}\bigg]+\log\phi^2,\label{V_eff}
\end{eqnarray}
where we have redefined the coupling constant by 
 \begin{eqnarray}
\frac{N_0^2}{2}{\alpha}^4=\tilde{\alpha}_0^4.
\end{eqnarray}
The difference between the result on ${\bf S}^2$ and this result lies in the replacement $\tilde{\alpha}\longrightarrow\tilde{\alpha}_0$ and the replacement $c_2\longrightarrow c_2^0$ in the definition of $\mu$. The analysis of the phase structure is therefore identical.

For example, the local minimum $\phi=\phi_0$ disappears for $\tilde{\alpha} < \tilde{\alpha}_*$. The critical curve $\tilde{\alpha}_*$ is determined from the point at which the real roots of $\partial V_{\rm eff}/ \partial \phi =0$ merge and disappear. This interpolates between $\tilde{\alpha}_*\sim N$ at small $M$ and the large $M$ result
\begin{eqnarray}
\tilde{\alpha}_*=3\big(\frac{2}{M}\big)^{1/4}.\label{pre2}
\end{eqnarray}
Thus, as the system is heated, the radius, $R$, expands 
form $R=1$, at large $\tilde{\alpha}$ to some critical value $R_*$ at $\tilde{\alpha}_*$.
When $\tilde{\alpha}<\tilde{\alpha}_*$ the fuzzy sphere solution no longer exists and the fuzzy four-sphere evaporates.

Furthermore, defining the entropy by ${\cal S}=<S>/N^2$, we obtain in the fuzzy four-sphere phase near the critical point the formula \cite{DelgadilloBlando:2008vi} 
\begin{eqnarray}
{\cal S}
&=&{\cal S}_*-\frac{24}{\phi_*\tilde{\alpha}_*^{\frac{5}{2}}\sqrt{M}}\sqrt{\tilde{\alpha}-\tilde{\alpha}_*}.\label{critentropy}
\end{eqnarray} 
This predicts immediately that the transition has a divergent 
specific heat with exponent $\alpha=1/2$, and also predicts that the entropy has a discrete jump, with a narrowing critical regime as $M$ is increased. However,  since the effective potential approximation does not take into account the coupling  $S_{12}$ between the two spheres, the value of the predicted discrete jump is not expected to agree with the Monte Carlo result. Nevertheless, we have shown by means of Monte Carlo \cite{Ydri_in_preparation} that the effective potential approximation remains   a very good fit to the Monte Carlo data especially for large values of $M$ where the coupling between the two spheres is dominated by the individual actions.


{\bf The phase diagram:}
In Monte Carlo simulations we use the Metropolis
algorithm and the action (\ref{main2}).
The errors were estimated using the  jackknife method.

\begin{figure}
\begin{center}
\includegraphics[width=8cm,angle=-0]{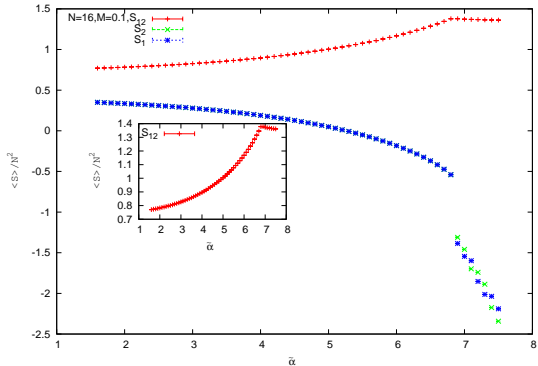}
\caption{The action for the $6$d Yang-Mills matrix model.}
\label{ac}
\end{center}
\end{figure}

The first estimation of the location of the transition is obtained from the intersection point of the average value of the action $<S>$ for different values of $N$. This intersection point is associated with a discrete jump in the entropy which is neatly observed for small values of $M$ (figure \ref{ac}). As $M$ increases it becomes harder to resolve the discontinuity.

\begin{figure}
\begin{center}
\includegraphics[width=8cm,angle=-0]{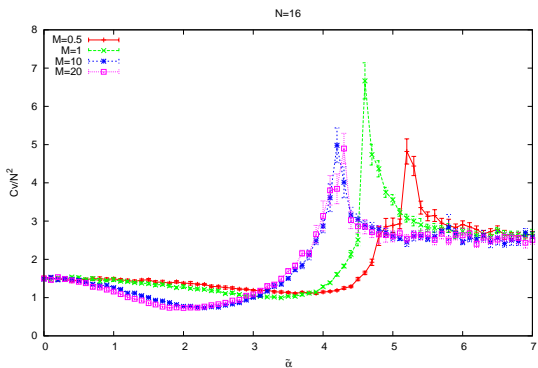}
\caption{The specific heat for the $6$d Yang-Mills matrix model.}
\label{Cvm0}
\end{center}
\end{figure}

For small values of $M$ (figure \ref{Cvm0}) a divergence in the specific heat,
$C_v:=<(S-<S>)^2>/N^2$, is observed. The maximum coincides with the intersection point of the action, and thus it marks the location of the transition. The theoretical prediction (\ref{pre2}) gives also a reasonable fit in this regime.

In summary, we have the behavior 
\begin{equation*}
   \frac{C_v}{N^2}  \longrightarrow \begin{cases}
                
                \frac{5}{2}~,~\tilde{\alpha}>>\tilde{\alpha}_*               & \text{fuzzy four-sphere phase}\\
               \frac{3}{2}~,~\tilde{\alpha}<<\tilde{\alpha}_*                & \text{Yang-Mills matrix phase.}\\
           \end{cases}
\end{equation*}
The location of the transition, for large values of $M$, moves to the minimum of the specific heat, and it agrees very well with the theoretical curve  (\ref{pre2}), while the intersection point of the action gives a lower estimate of the transition point in this case. 

The maximum of $C_v$, for large values of $M$, saturates around the value $\tilde{\alpha}\sim 4.2$. Indeed, starting from some value of $M$ around $M\sim 1$, the peak in $C_v$ occurs always at this value $\tilde{\alpha}\sim 4.2$ . This is the regime where the transition from the fuzzy four-sphere phase to the Yang-Mills matrix phase becomes a crossover transition. 
The critical line between the fuzzy four-sphere phase and the crossover phase is given by the maximum of $C_v$, whereas the critical line between the Yang-Mills matrix phase and the crossover phase is given by the minimum of $C_v$.


As the value of $M$ is increased, our numerical study confirms that 
the fuzzy four-sphere to matrix model transition is shifted to
lower values of $\tilde{\alpha}$, and extrapolating $M\rightarrow\infty$ 
we infer that the critical coupling goes to zero. In other words, the fuzzy four-sphere phase is only stable in the limit
$M\longrightarrow\infty$.

\begin{figure}
\begin{center}
\includegraphics[width=8cm,angle=-0]{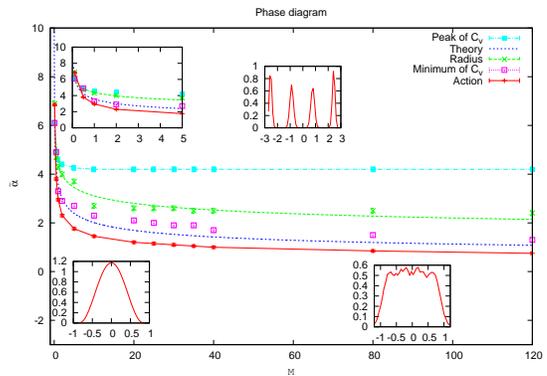}
\caption{The phase diagram of the $6$d Yang-Mills matrix model.} 
\label{phasediag}
\end{center}
\end{figure}

Our results are summarised in a phase diagram in figure \ref{phasediag} which also include measurement from the radius \cite{Ydri_in_preparation}. As in the $2-$dimensional case studied in \cite{DelgadilloBlando:2007vx}, the persistence of the critical line, as determined by the crossing point of the average action at the minimum of $C_v$, suggests that the transition is $2$nd order. This is consistent with the theoretical analysis (\ref{critentropy}) which indicates a divergent specific heat with exponent $\alpha=1/2$ but with a narrowing critical regime as $M$ is increased. See also \cite{O'Connor:2013rla}. However, for large values of $M$ the behavior seems to be quite different with the appearance of a crossover phase separating the fuzzy four-sphere phase from the Yang-Mills matrix phase.

{\bf The eigenvalue distributions:}

\begin{figure}[htbp!]
\begin{center}
\includegraphics[width=8cm,angle=-0]{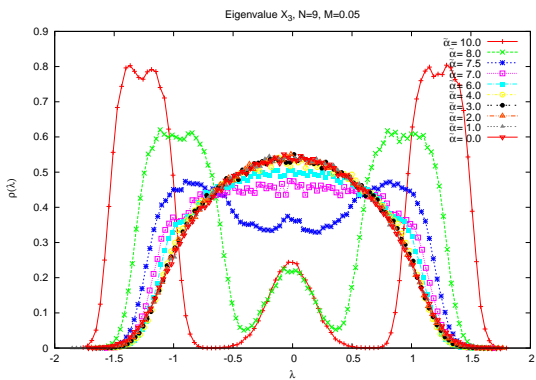}
\caption{The eigenvalue distribution for $X_3$ across the transition line. }
\label{D3fuzzy}
\end{center}
\end{figure}

\begin{figure}[htbp!]
\begin{center}
\includegraphics[width=8cm,angle=-0]{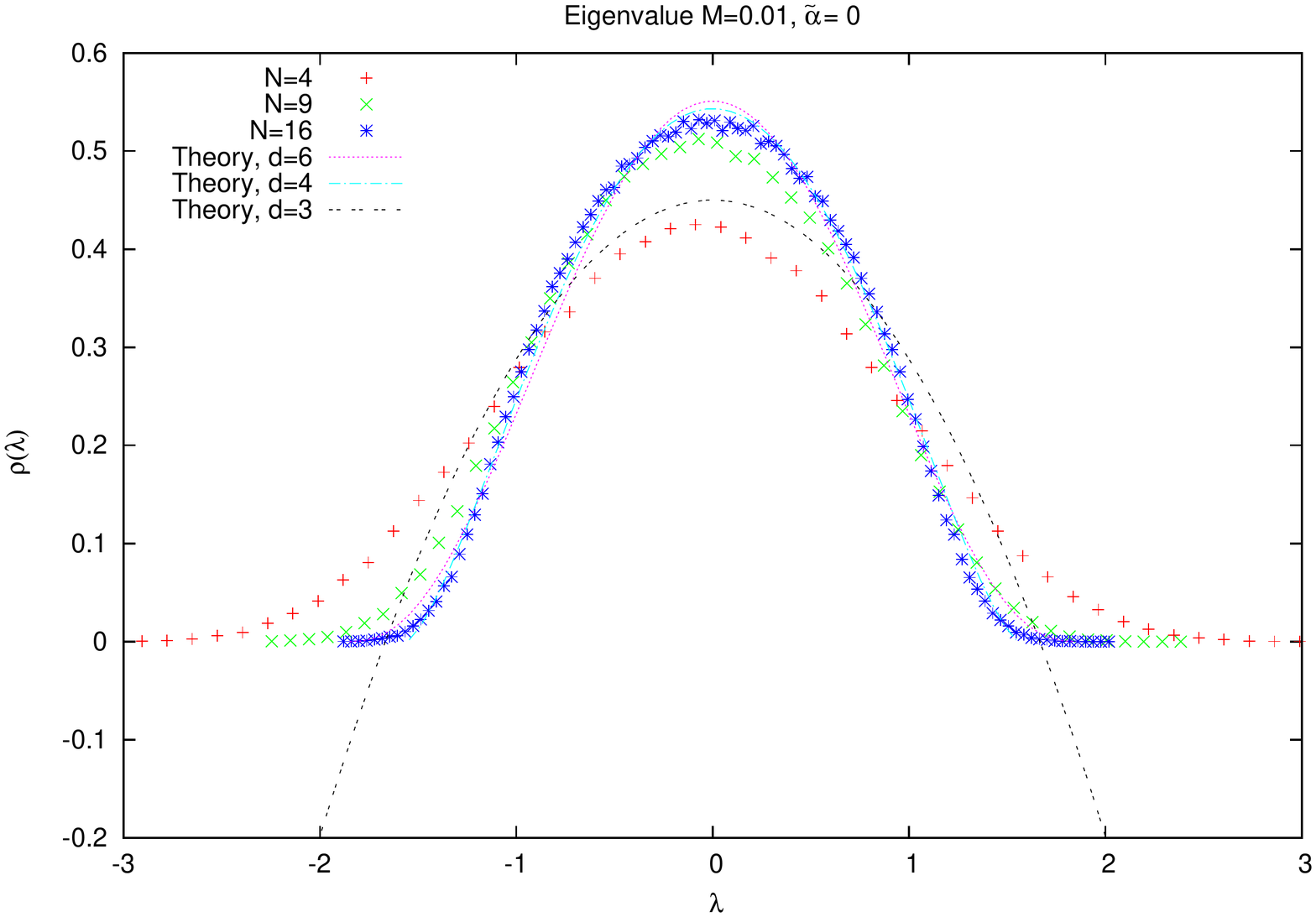}
\caption{The eigenvalue distributions for $X_3$ for small values of $M$ and $\tilde{\alpha}$.}
\label{D3matrix}
\end{center}
\end{figure}

The most detailed order parameter at our disposal is the distribution of eigenvalues of observables.  Here, we focus mainly on $X_3$ and $Y_3$. The characteristic behaviour of the distributions of 
eigenvalues in the fuzzy four-sphere and Yang-Mills matrix phases is illustrated in figures \ref{D3fuzzy}
and \ref{D3matrix} respectively.

For small values of $M$, we see that, as one crosses the critical curve 
in figure \ref{phasediag}, the eigenvalue distribution of $X_3$ and $Y_3$ undergoes a 
transition from a point spectrum given by the eigenvalues of the $SU(2)$ generators in the largest irreducible representation which is of size $N$, viz 
\begin{eqnarray}
+\frac{N-1}{2},\frac{N-1}{2}-1,...,-\frac{N-1}{2}+1,-\frac{N-1}{2},\label{point}
\end{eqnarray}
to a continuous distribution
symmetric around zero given by the $d=6$ law \cite{Ydri:2012bq, Filev:2013pza, O'Connor:2012vr, Filev:2014jxa}.
\begin{eqnarray}
\rho(\lambda)=\frac{\Omega_{d-1}}{V_d(d-1)}(r^2-\lambda^2)^{(d-1)/2},\label{d6}
\end{eqnarray}
This is a generalization of the $d=3$ (parabolic) law found in $2$ dimensions \cite{Berenstein:2008eg,DelgadilloBlando:2012xg}. This can be derived from the assumption that the six matrices are commuting with a joint eigenvalue distribution uniform inside a $6-$dimensional ball with a radius $r$.

\begin{figure}[htbp!]
\begin{center}
\includegraphics[width=8cm,angle=-0]{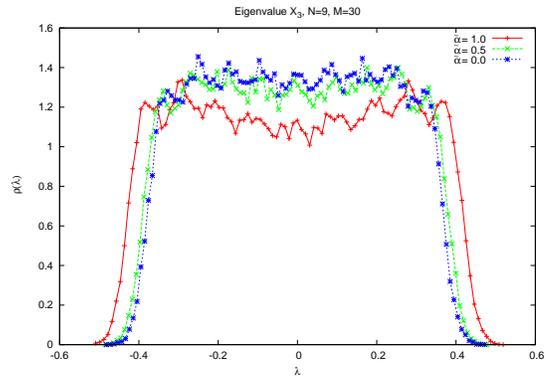}
\caption{The eigenvalue distributions for $X_3$ for large values of $M$ and small values $\tilde{\alpha}$.}
\label{D3uniform}
\end{center}
\end{figure}

However, for large values of $M$ the behavior of the distribution inside the Yang-Mills matrix phase changes to a uniform distribution. See figure (\ref{D3uniform}). This occurs in the regime of the crossover phase. Indeed, for large value of $M$, in the crossover phase, a strong gauge field is superimposed on the fuzzy four-sphere background in such a way that the middle peaks flatten then disappears
slowly in favor of a uniform distribution. The last peaks to go are the maximum and the minimum of the $SU(2)$ configuration (\ref{point}).


{\bf Conclusions:} 

In this letter we have extended our previous work \cite{DelgadilloBlando:2007vx} to $4$ dimensions. 
We studied a six matrix model with global $SO(3)\times SO(3)$ symmetry containing
at most quartic powers of the matrices proposed in \cite{DelgadilloBlando:2006dp}. The value $M=1/2$ of the deformation corresponds to the model of \cite{Behr:2005wp}. This theory exhibits a phase transition from a geometrical phase at low temperature, given by a fuzzy four-sphere ${\bf S}^2_N\times{\bf S}^2_N$ background, to a Yang-Mills  matrix phase with no background geometrical structure at high temperature. The geometry as well as an Abelian gauge field and two scalar fields are determined dynamically as the temperature is decreases and the fuzzy four-sphere condenses. The transition is exotic in the sense that we observe, for small values of $M$, a discontinuous jump in the entropy, characteristic of a 1st order
transition, yet with divergent critical fluctuations and a divergent
specific heat with critical exponent $\alpha=1/2$. The critical temperature is pushed upwards as the 
scalar field mass is increased (see figure \ref{phasediag}).  For small $M$, the system in the Yang-Mills phase is well
approximated by $6$ decoupled matrices with a joint eigenvalue distribution which is uniform inside a ball in ${\bf R}^6$. This yields the $d=6$ law (\ref{d6}). For large $M$, the transition from the four-sphere phase to the Yang-Mills matrix phase turns into a crossover and the eigenvalue distribution in the Yang-Mills matrix phase changes from the $d=6$ law to a uniform distribution. 

 In the Yang-Mills matrix phase the specific heat is equal to $3/2$ which coincides with the specific heat of $6$ independent matrix models with quartic potential in the high temperature limit  and is therefore consistent with this interpretation. Once the geometrical phase is well established the specific heat 
takes the value $5/2$ with the gauge field contributing $1/2$ \cite{Gross:1980he} and the two scalar fields each contributing $1$ \footnote{Recall that in the $3$d Yang-Mills matrix model the specific heat takes the value $1$ in the geometrical phase which is attributed in this case to the normal scalar field since there is no propagating gauge degrees of freedom in $2$ dimensions.}. Therefore, the role of dark energy in this model is played by the two scalar fields, which are fully decoupled from the gauge field at large $M$, yet they contribute $80$ per cent of the total specific heat of the theory.

The physical radius of the two spheres ${\cal R}$ which is  defined by (\ref{radiuss}) is behavior is such that it goes to a minimum value ${\cal R}_{\rm min}$, which can be computed using the  $d=6$ law (\ref{d6}) for small values $M$, in the Yang-Mills matrix phase, while in the fuzzy four-sphere it increases for large $\tilde{\alpha}$ as $\tilde{\alpha}^2$, i.e. the radius expands with the temperature as $1/\sqrt{T}$. 

The model presents thus an
appealing picture of a geometrical phase emerging as the system cools
and suggests a scenario for the emergence of geometry in the early
universe.

\paragraph{Acknowledgments:}
This research was supported by CNEPRU: "The National (Algerian) Commission for the Evaluation of
University Research Projects"  under contract number ${\rm DO} 11 20 13 00 09$.

\end{document}